**Pre-charging polymer surfaces enhances droplet mobility and electrification**

*Shuaijia Chen[1], Kenta Morita[2,3], Dumindu Dassanayaka[4], Hans-Jürgen Butt[5], Peter C. Sherrell[1,2],* Amanda V. Ellis[1],* Joseph D. Berry[1]**

[1]Department of Chemical Engineering, The University of Melbourne, Parkville, 3010, Victoria Australia

[2]Department of Chemical Science and Engineering, Graduate School of Engineering, Kobe University 1-1 Rokkodai, Nada-ku, Kobe 657-8501, Japan

[3]Research Center for Membrane and Film Technology, Kobe University, 1-1 Rokkodai, Nada-ku, Kobe 657-8501, Japan

[4]School of Science, RMIT University, Melbourne, 3000, Victoria, Australia

[5]Max Planck Institute for Polymer Research, Ackermannweg 10, 55128 Mainz, Germany

E-mail:

peter.sherrell@rmit.edu.au; amanda.ellis@unimelb.edu.au; berryj@unimelb.edu.au

*Abstract*

Surface-bound electric charge on polymer materials can strongly influence droplet behaviour and solid–liquid charge transfer, but the mechanisms and the means to control these effects remain unclear. In this work, we systematically controlled the surface charge on polymer surfaces, including polytetrafluoroethylene (PTFE) and Nylon-66, by first neutralising the surfaces with an anti-static ion blower and then applying charge using an ion gun. We find that droplets pick up pre-deposited surface ions during the first wetting of the surface, and that the transferred charge directly correlates with the deposited charge encountered by the wetted area for moderate deposited densities ($|\sigma_d| \lesssim 40\ \mu\text{C/m}^2$) independent of material properties. We also demonstrate that the deposited charge reduces contact angle and increases contact-line mobility in a manner consistent with an increase in effective solid surface energy. For higher surface charge densities, we observe instabilities such as droplet splitting or detachment. This work demonstrates an effective approach to control solid–liquid electrification, enabling amplification or suppression of surface charge and the directed manipulation of fluid motion on surfaces.

## 1. Introduction

The electrification of water moving over surfaces has been studied for over a century, since Lenard and Thomson [1, 2] both observed an electrical effect when a liquid drop struck a liquid-film-coated plate. Such electrification phenomenon was attributed to the formation of a "double coating of electricity", where the magnitude and sign of electricity are highly dependent upon the nature of the surrounding gas [1, 2]. Building upon the discovery of Lenard and Thomson, the charge separation model was introduced to explain the charge transfer that occurs when a droplet slides down a surface (the "slide electrification" phenomenon). This model involves the spontaneous formation and separation of the electric double layer (EDL) in the vicinity of the receding contact line through the spontaneous deposition of ions (typically hydroxide ions) or the dissociation of surface groups on the dewetted surface [3-14], where the bound surface charges (within the adsorption layer) are deposited at the edge of the receding contact line and adsorbed by the dewetted hydrophobic surface, causing the surface to become electrically charged. The counter charges (within the diffuse layer) accumulate within the droplet [15-20].

Electrification *via* charge separation is determined by several factors, including the physical properties of the solid surfaces (hydrophobicity), contact line velocity, and the ionic concentration of the liquid droplet. Ratschow et al.[11] demonstrated that charge separation is more significant on hydrophobic surfaces, and the amount of separated charge decreases when the droplet velocity exceeds 1 cm $s^{-1}$. Further, the bound surface charge separation is governed by the local EDL structure near the contact line, which is influenced by the gas-liquid interface and the internal flow of the liquid [11]. Li et al.[21] demonstrated that the addition of ionic or non-ionic surfactants can reduce charge transfer due to the screening effect of both ionic and non-ionic surfactants. Helseth [22] demonstrated that the presence of cationic surfactant effectively quenches the negative charges in a single stage, whereas anionic (SDS) and neutral (Triton X-100) surfactants exhibit a complicated two-stage charge transfer response, where charge transfer is inversely proportional to the surfactant concentration. Furthermore, Zhou, X., et al.[23] showed that the slide electrification on dielectric surfaces has a selectivity for depositing dissolved solutes, where deposition is strongly affected by the solute polarity and drop charge. The deposition of charged solutes occurs at the receding contact line, where the electric field generated from slide electrification at the rear side of the droplet leads to a preferential deposition of the charged solutes.

Many research teams have continued to investigate slide electrification, however, building a holistic mechanistic picture remains a major challenge, due to the interrelation of key governing factors. For example, the electrostatic force from slide electrification significantly affects the droplet motion and contact angle [24-26]. This effect arises due to the electrostatic interaction between a drop and the presence of surface charges deposited by previous droplets [11, 24, 25, 27]. Slide electrification is also complicated by contact angle hysteresis, where the contact angle varies locally in the range between receding and advancing contact angles as the droplet slides. Electrostatic interaction between surface charges and counter charges within the droplet increases the energy requirement to dewet surfaces [14]. Yatsuzuka et al.[13] suggested that the correlation between the dielectric layer thickness and the Coulombic force exerted from interfacial charges plays an important role in contact electrification. Hinduja[28] demonstrated that the charge separation is proportional to the width of the droplet instead of the length of the receding contact line and the droplet's width variation is attributed to the force magnitude difference, which alters the friction force and torque. Hence, the electrification of a liquid drop is similar to the friction force exerted on the droplet [29].

An alternative approach to study droplet electrification is the sessile drop method, which allows for the decoupling of wetting and dewetting effects [30, 31]. Previously, Chen et al. demonstrated that pinning-depinning motion of the contact line during wetting also causes charge transfer, resulting in significantly enhanced electrification [30]. They also observed that charge transfer can occur during first wetting, even without pinning-depinning occurring, possibly due to the surface treatment method used which involved wiping the polymer surface with a Kimwipe. However, the effects of surface history due to different surface treatment methods were not fully investigated. We believe that the surface treatment method may lead to the introduction of surface charge which may have substantial effects on charge transfer during wetting. Most studies relevant to liquid-solid charge transfer use neutralised surfaces [11, 24, 25], whereas some studies use slide electrification to charge the surface [32, 33], and others use solid-solid pre-charging to promote the pinning-depinning of contact line motion [30]. Herein, we aim to address these key knowledge gaps by proposing a hypothesis that pre-existing surface charge present on the dry polymer surface plays a key role in charge transfer during the first wetting stage.

Herein, we have studied the role of surface charge, controlled by an ion blower or ion gun, on droplet deposition and subsequent electrification. Through the combination of a sessile-drop goniometer and charge measurement, the electrification phenomena and charge generation mechanism within the solid-liquid interface can be probed at different surface charge densities ranging from negative (-60 $\mu C/m^2$) to positive (+60 $\mu C/m^2$). The effects of surface morphology, hydrophobicity, hydrophilicity, chemical properties (e.g. dielectric and electronegativity) and different positions on the triboelectric series on contact line motion and electrification within the solid-liquid interface have been investigated. Therefore we used different polymer surfaces, namely polytetrafluoroethylene (PTFE) and polyamide (nylon). [34].

## 2. Methods

### *2.1. Sample Fabrication*

Samples were fabricated with commercial polymer including polytetrafluoroethylene (PTFE), (dielectric permittivity 2.1, Swift Supplies, Australia) and poly[imino(1,6-dioxohexamethylene) iminohexamethylene] (Nylon 6,6) (dielectric permittivity 3.3, Goodfellow, China) with a dimension of 3×3 $cm^2$ and thickness of ~500 ± 2 µm. A 3 nm chromium (Cr) layer followed by a 30 nm gold (Au) layer was sputtered on one side of each polymer sheet as a conductive electrode (Quorum Q150T ES, Quorumtech, UK). A 15 cm electrical wire (FLEXI-E 0.15, Stäubli Electrical Connectors AG) was soldered (solder: 0.3 mm lead-free Weller Wire, RS Components, Australia; soldering iron: JBC-CD-2SQE, Mektronics, Australia) onto the non-adhesive side of copper (Cu) foil tape (70 µm thick, 1181, 3M, USA). The adhesive conductive side of Cu foil tape was then adhered to the sputter-coated electrode side of the polymer sheets. The electrode side (single electrode mode, no liquid drop contact with) of each polymer sample was insulated and protected with Kapton (polyimide) tape (50 µm thick, RS Components, Australia). To ensure the contact surface was flat and the testing samples were swappable after each trial, double-sided tape (3M) was used to adhere the encapsulated electrode side of polymer sample with a glass substrate (Premiere, Microscope Slides, 2” × 3” 1.00 mm) (Figure S1).

### *2.2.Pre-charging Surface Treatment*

Prior to the liquid droplet measurement, the polymer sample was placed onto the sample stage (insulated with Kapton tape). Surfaces were neutralised using an ionizing air blower (IBT6432E Simco-Ion Ionizing Blower) (Figure 1 a). The polymer surface was blown for 5 minutes to ensure the surface was fully neutralised and the effects from residual surface charge were eliminated [11, 24, 30, 35]. To manipulate the magnitude and polarity of the initial surface charge on the polymer sample, the surface was pre-charged via ionised air generated from the piezo-electric crystal within the anti-static gun (Milty Zerostat 3 Anti-Static Gun) (Figure 1 b). The pre-charging polarity was controlled via the direction of the trigger action to either pull (positive) or release (negative). The magnitude of pre-charging was adjusted by the number of trigger actions. During pre-charging, the polymer sample's electrode was connected to the electrometer to monitor the magnitude and sign of surface charge. The electrometer reports the total charge on the electrode at the back of the polymer within the 0 - 2 µC (neutralising and pre-charging) and 0 - 20 nC (droplet testing) range, providing a nominal resolution of 1 pC and 10 fC, respectively.

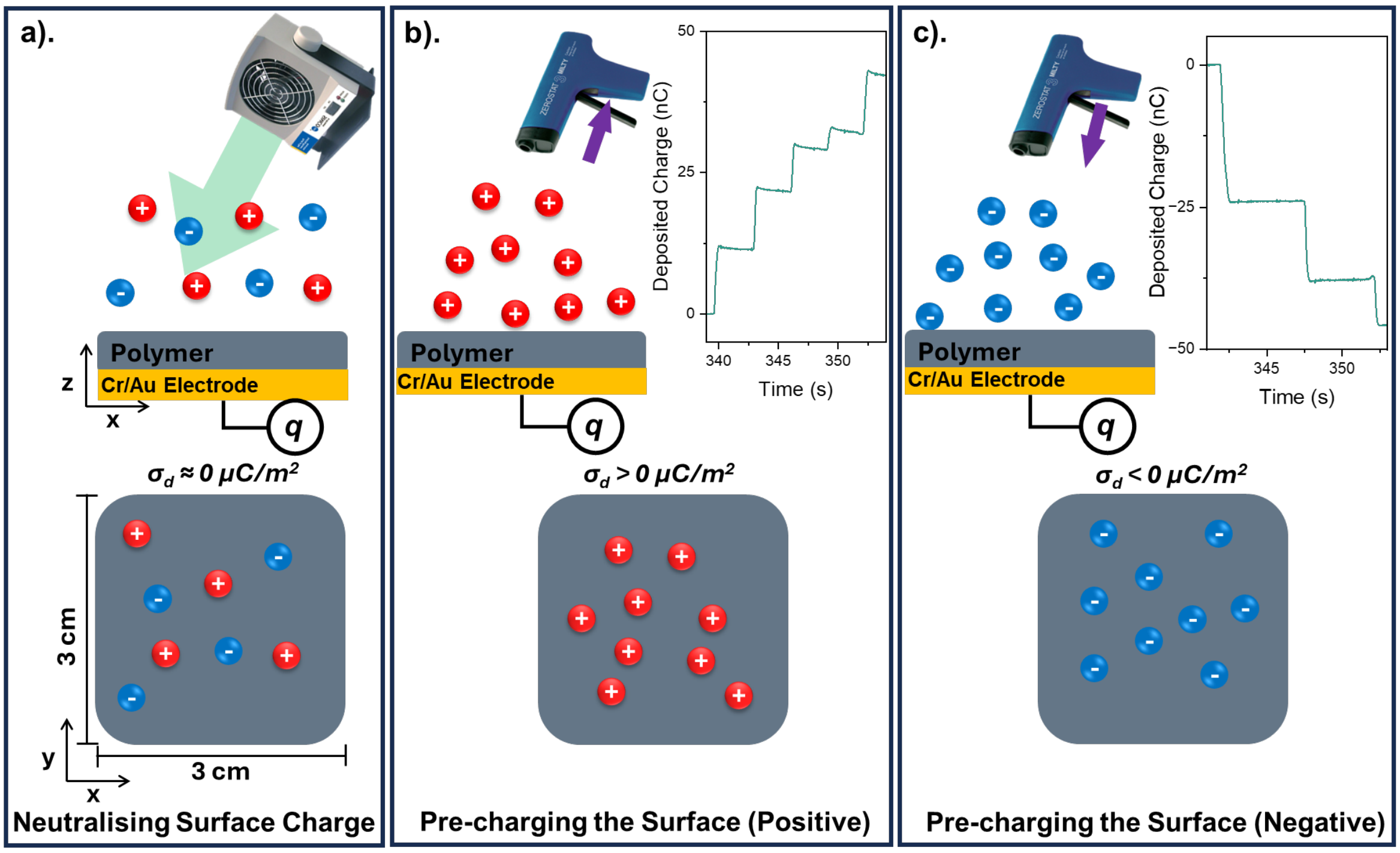

**Figure 1. a.** Schematic of neutralising surface charge with the ionizing blower to eliminate the effect of residual surface charge on the tested polymer surfaces. b. Schematic of injecting charge to the testing surface with an anti-static gun to pre-charge the surface prior to the measurement, and example of surface charge measurement with surface pre-charging protocol within the positive regime. **c.** The negative regime

When charging the surface using the ion gun, we found that the electrometer measured significant charge leakage possibly due to ions landing directly onto the electrode or the wires connecting the electrode to the electrometer. To circumvent this, we used a 3d printed shield (Figure S2) to direct ions onto the surface. We also used an electrostatic field meter (SIMCO FMX-004, accuracy +/- 10%), which contains an oscillating voltage sensor to independently measure the surface potential as a function of deposited surface charge. We found that the shield prevented some charge leakage, however it did not prevent all of it as the electrometer continued to show an increase in surface charge even when the field meter showed a constant surface potential.

To measure the deposited surface charge, we developed a two-step protocol to calibrate the electrometer reading. In the first step we charged the polymer surface by rubbing it with a Kimwipe, which introduces negligible charge leakage. We measured the charge transfer using the electrometer, as well as the surface potential using the field meter. We then converted the field meter surface potential $U$ to a surface charge density using the relationship $\sigma = \varepsilon_0 \varepsilon_r U/d$, where $\varepsilon_0$ is the permittivity of free space, $\varepsilon_r$ is the relative permittivity of the polymer and $d$ is the thickness. This relationship assumes an infinite flat plate, a homogeneous charge distribution and it introduces some error given the sample size of $3 \times 3 \pm 0.01$ cm$^2$. The corresponding electrometer and field meter measurements were then plotted (Figure S2 c), allowing us to calibrate the field-meter measurements assuming that the electrometer measurements were correct. We found that the surface charge density deposited via rubbing could be estimated from the field-meter surface charge density estimate as $\sigma_d = 1.7\ \sigma_{FM}$. In the next step, we charged the surface using the ion gun, and again recorded the surface charge density estimates from both the electrometer and the field-meter. We then corrected the field-meter data using the equation above, and plotted the corrected field-meter surface charge density against the electrometer surface charge density $\sigma_E$ (Figure S2 d). We then fitted a function to allow us to estimate the actual deposited surface charge density from the electrometer reading. The function that fitted the data best was:

$$\sigma_d = 55.6\ \tanh(0.011\sigma_E)$$

This relationship was then used to give us the deposited charge density on the polymer surface.

For subsequent testing, (see [30] for more details on the procedure) a programmable syringe pump (NE-1000, NewEra Pump System, USA) was adopted to manipulate wetting and dewetting of the surface by adjusting a droplet of deionised water (Milli-Q, resistivity 18.2 MΩ·cm) of volume ($V$) via the volumetric flow-rate ($Q$). Each experimental trial consisted of 2 cyclic wetting (dispensing) and dewetting (withdrawing) stages (more cyclic testing results are included in Figure S3), with a constant change of droplet volume ($\Delta V$) of 150 µL under a fixed $Q$ (1200 µL/min). To differentiate the effects of wetting and dewetting, we paused 10 s in between each stage. A CMOS camera (Basler acA1920-150m, BASLER, Germany) was used to capture the contact line motion and shape of liquid drop at 50 fps. An electrometer (KEITHLEY-6514, Tektronix, USA) connected with a data acquisition system (DAQ, NI-9223, National Instruments, USA) was used to measure electrical charges. Contact angle, contact point position, contact radius, and contact area data were extracted from recorded videos using custom Python code and ImageJ. The electrical measurements were performed within the range of [-2000, 2000] nC (neutralising & pre-charging) and [- 20, 20] nC (droplet contact). The range and resolution for pre-charging and droplet testing stages were maintained within a suitable measurement range to achieve consistent and reproducible measurements. All the experiments and measurements were performed under ambient conditions with a room temperature of 25 ± 3 ̊C and a relative humidity of 50 ± 10%. According to relevant literature, the effects of humidity on charge separation, accumulation, and discharge are negligible within the range of 10-70% [14, 23, 36]. To evaluate whether the metallic needle plays a role in charge transfer during wetting and dewetting, tests were performed with the needle directly grounded via an electric wire and compared to tests without directly grounding the needle. The results show that the needle has no significant effect on charge transfer (Figure S4).

## 3. Result and Discussion

### *3.1. Effect of Surface Charge on Charge Transfer*

To evaluate the effect of surface pre-charging, charge measurements were performed over the first two wetting/dewetting stages (WS/DS, Figure 2a). The measured charge signal during the wetting/dewetting stages on both PTFE and Nylon surfaces is shown in Figures 2 b-g for different deposited charge densities. When the liquid begins to infuse into the initial sessile drop, the droplet grows, and the contact line starts advancing (Figure 2 a stage 1). Depending upon the magnitude and polarity of the initially deposited charge, charge transfer is observed as wetting proceeds. For the ionising air blower neutralised PTFE (Figure 2 b) and Nylon (Figure 2 c) surfaces, there is no charge transfer ($\Delta q \approx 0$) during the first wetting stage (“1st WS”, red region, Figure 2 b-g). The first observed charge transfer on neutralised surfaces occurs on the subsequent first dewetting stage (“1st DS”, yellow region, Figure 2 b-g). It matches the theory of the charge separation model, in which electrical charge is deposited during dewetting, causing the surface to become electrically charged [14, 37].

These surface charges deposited by the droplet are picked up again on the second wetting stage (blue region, Figure 2 b-g).

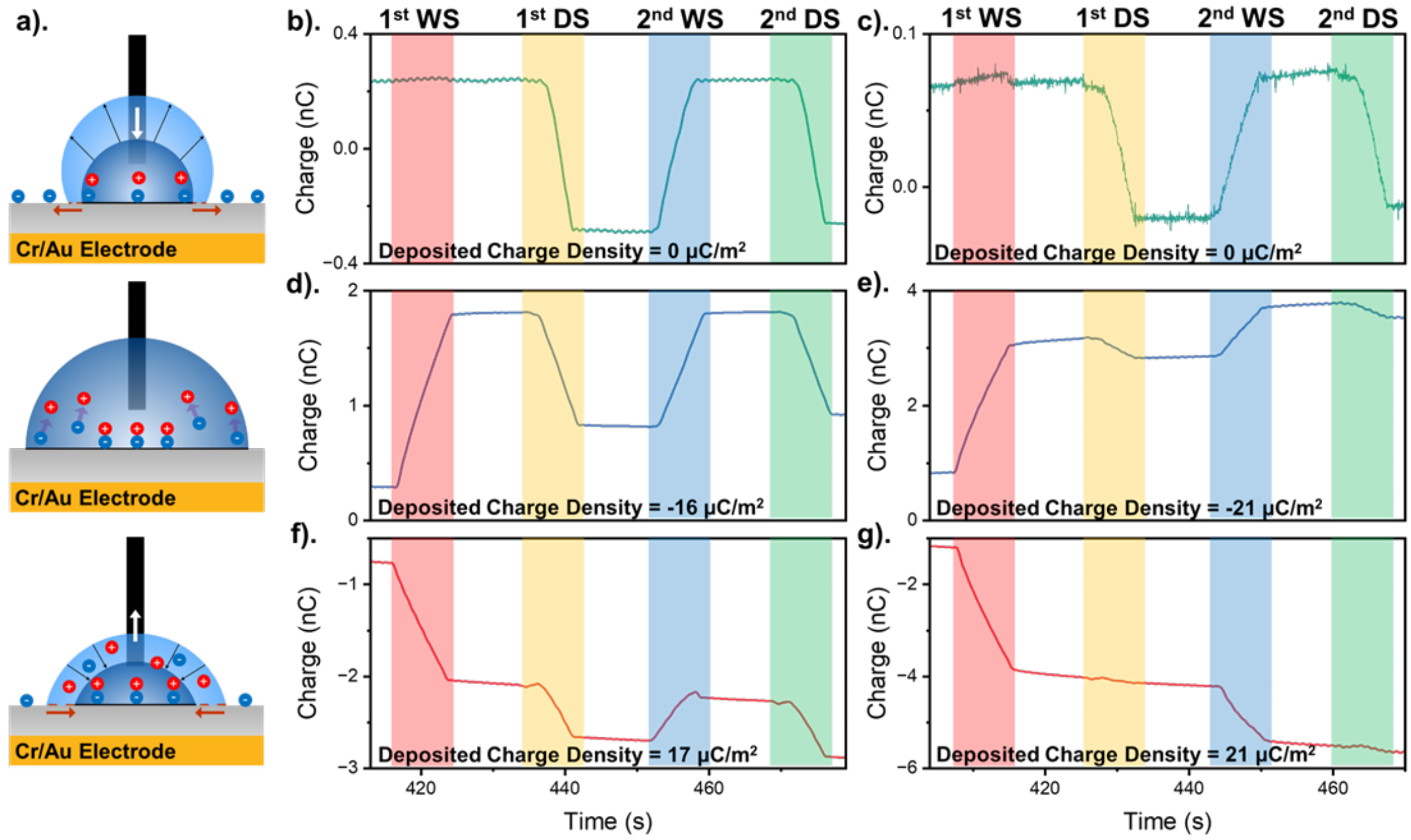

**Figure 2. a**. Schematic diagram of droplet interaction with deposited surface charge on dielectric surface during wetting (WS), pause, and dewetting (DS) stages on negatively pre-charged surface; corresponding charge measurement of the first two wetting and dewetting stages **b.** on the ionizing air blower neutralised PTFE surface **c.** Nylon surface; **d.** the anti-static gun negatively charged PTFE surface; **e**. Nylon surface; **f.** the anti-static gun positively charged PTFE surface; **g.** Nylon surface; The shaded regions distinguish individual wetting and dewetting stages, and low pass filter with a cutoff frequency at 1 was applied to remove background noise.

In contrast, for positive or negative pre-charged surfaces, charge transfer from the surface to the liquid occurs upon first wetting ((“1$^{st}$ WS”, red region, Figure 2, b-g). The charge transferred was of the opposite sign as the deposited surface charge. The negatively pre-charged protocol, with -16 μC m$^{-2}$ and -21 μC m$^{-2}$ initial deposited surface charge density for PTFE and Nylon (Figure 2 d and e), respectively, shows a large charge transfer ($\Delta q$) during the first wetting stage on PTFE (~1.5 nC) and Nylon (~2.2 nC) surfaces. The positively pre-charged protocol (Figure 2 f and g) also shows large charge transfer during the first wetting stage on PTFE (~1.2 nC) and Nylon (~2.7 nC) surfaces. These observations suggest that electrical charges present on the substrate due to exposure to an ion-gun are taken up by the droplet during wetting, leading to the direct correlation with measured surface charge.

The magnitude of $\Delta q$ from the remaining wetting and dewetting stages being higher than $\Delta q$ observed on neutralised PTFE (~0.52 nC) and Nylon (~0.08 nC) surfaces also supports the mechanism of surface charge taken up by droplet, suggesting that the greater presence of unbalanced charge available to move in the system. Less charge transfer is observed on Nylon during dewetting than on PTFE, because the drop exhibits larger contact angle hysteresis and the contact line remains pinned for longer rather than moving across the surface. As a result, the magnitude of $\Delta q$ observed during dewetting on Nylon was small.

To further investigate the influence of the magnitude and polarity of surface charge on solid-liquid charge transfer, over 500 charge measurements were performed with deposited surface charge density $\sigma_d$ within the range of $\pm$ 60 μC/m$^2$ on both PTFE and Nylon surfaces. Figure 3 shows the correlation between $\Delta q$ and the deposited surface charge density $\sigma_d$ for different wetting and dewetting stages. A linear relationship was observed between $\Delta q$ and the initial deposited surface charge density during the first wetting process (Figure 3 a). For small deposited surface charge, within the $\pm$ 40 μC m$^{-2}$ range, the charge transfer $\Delta q$ during wetting increases linearly on both PTFE (slope of -0.078 nC per μC/m$^2$) and Nylon (slope of -0.1 nC per μC/m$^2$) surfaces. However, for larger deposited surface charge densities $|\sigma_d| \geq 40$ μC/m$^2$, the charge transferred begins to show deviation from linearity on both PTFE and Nylon surfaces.

This increased variability of the response at $|\sigma_d| \geq 40$ μC m$^{-2}$ was supported by optical analysis of the deposited droplets, which showed a wide array of anomalous effects including increased droplet movement and droplet splitting (Figure 3 a, red inset; Figure S5) at $|\sigma_d| \geq$ 35 μC m$^{-2}$. The droplet tended to randomly split into multiple smaller drops on a highly positively charged PTFE surfaces (Figure 3 a, red box), whereas the droplet tended to split into a tiny and a larger drop on highly negatively charged PTFE surfaces (Figure 3 a, magenta box). For highly charged Nylon surfaces, the droplet tended to pin on the surface and detach from the needle during the dewetting stage (Figure 3b, brown box).

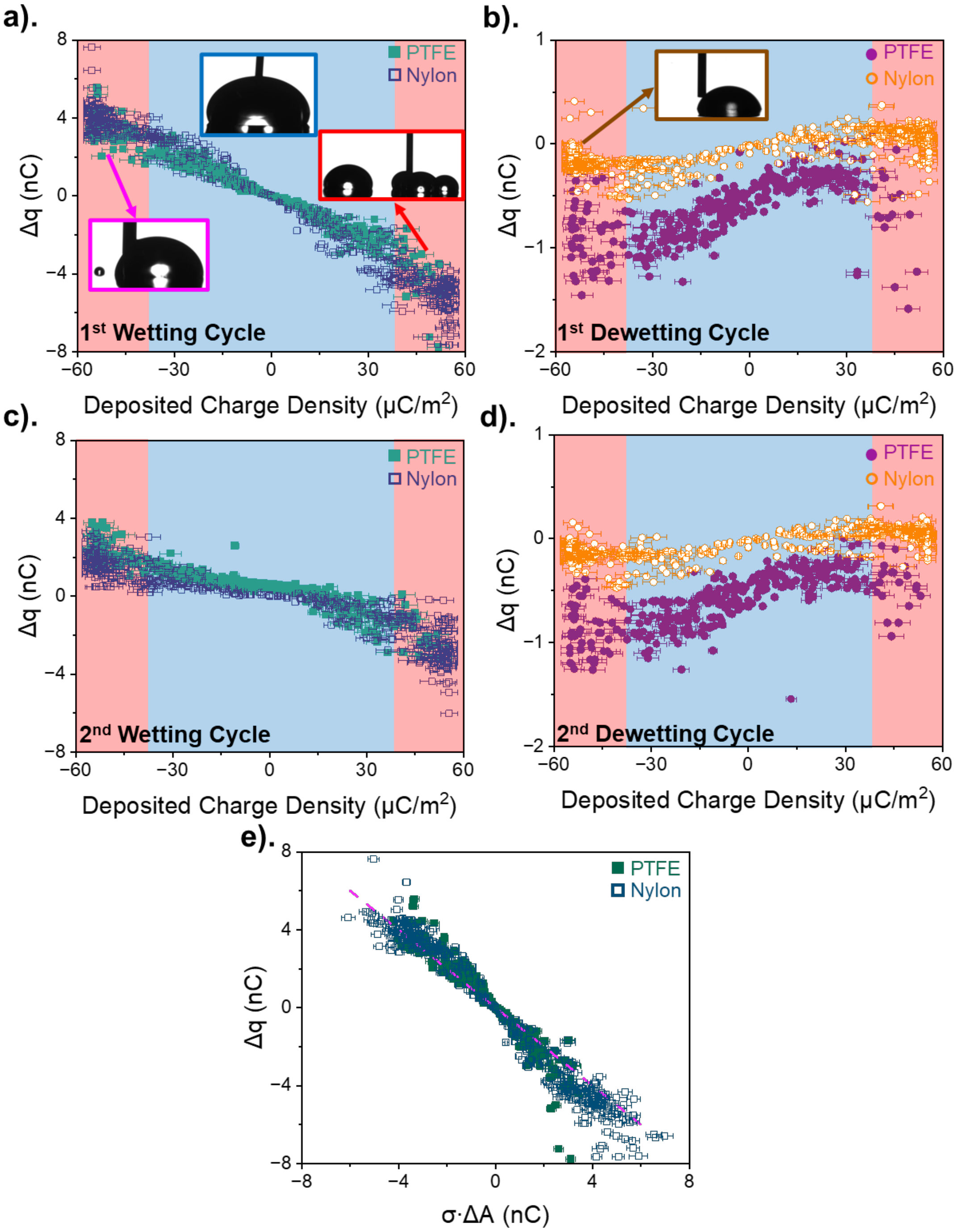

**Figure 3.** Charge transfer plotted against the magnitude of deposited surface charge density on PTFE and Nylon surfaces during **a.** the first wetting cycle and examples of droplet images showing stable and split behaviours on the PTFE surface within the corresponding pre-charged region (red box for positive regime, blue box for neutral regime, and magenta box for negative); **b.** first dewetting cycle and example of detached droplet on Nylon surface within the corresponding pre-charged region (brown box); **c.** second wetting cycle; **d.** second dewetting cycle.; **e.** Change in charge Δq against the deposited surface charge encountered by the droplet as it first wets the surface (defined as the deposited surface charge density multiplied by the change of contact area during the first wetting stage) for a 170 μL droplet on **e.** PTFE (green solid box) and Nylon (blue outline box) surface. The dashed line represents a 1:1 relationship. Note that only the data from experiments with a single, stable droplet has been plotted in all subfigures.

To evaluate the critical value of charge density causing droplet splitting, we can relate the surface charge to drop stability via Rayleigh's criterion [38], yielding (see SI for details):

$$\sigma_s < \left(\frac{48X\varepsilon_0\gamma}{\pi}\left(\frac{2\pi}{3}\right)^{\frac{4}{3}}\right)^{\frac{1}{2}} V^{-\frac{1}{6}} \qquad \text{Equation 1}$$

where $V$ is the drop volume and $X$ is the fissility threshold, which is unity for a levitated droplet. Notably, no droplet splitting due to charge has been reported for sessile drops on non-lubricated surfaces, possibly due to contact line pinning [39]. However, recent work by Daniel et al. [39] has shown breakup due to charge occurs for sessile droplets on a lubricated surface (which inhibits contact-line pinning). The fissility threshold is $X = 0.25$ for this case. Using the same value of $X$ in Equation 2 gives a critical surface charge density of ~35 μC/m$^2$ for a 150 μL droplet, which in good agreement with our observations.

A similar relationship between the magnitude of deposited surface charge density and $\Delta q$ is present during the second wetting cycle on both PTFE (slope of -0.036 nC per μC/m$^2$ = 36 mm$^2$) and Nylon (slope of -0.032 nC per μC/m$^2$) surfaces (Figure 3 c), except that the charge transfer magnitude has decreased significantly due to there being less surface charge present, with most transfer occurring during the initial wetting stage.

For the dewetting stage of both first and second cycles on PTFE and Nylon surfaces (Figure 3 b and d), the polarity of deposited surface charge causes a distinct difference in the charge transfer. For the negative regime of deposited surface charge density of both cycles, the charge transferred $\Delta q$ during dewetting shows an approximately linear relationship (PTFE slope of 0.013 nC per μC/m$^2$ = 13 mm$^2$ and Nylon slope of 0.0065 nC per μC/m$^2$ = 6.5 mm$^2$) with the magnitude of deposited surface charge density within the range of ±40 μC/m$^2$; but reduces slightly with increased scatter at higher surface charge density (-40 to -60 μC/m$^2$ and +40 to +60 μC/m$^2$). The sign of $\Delta q$ during dewetting on the Nylon surface is consistent with the polarity of surface charge (positive in the positive regime, and negative in the negative regime) and there is polarity conversion across the negative and positive regimes of surface charge, where the sign of $\Delta q$ on the PTFE surface is negative across the entire range of deposited surface charge density (Figure 3 b and d). This observation could be attributed to the material's position on the triboelectric series based on the empirical experiments, where PTFE has a better affinity for anions, and Nylon has a better affinity for cations [34]. The comparison of change of charge density between wetting and dewetting for the same material and same cycle is plotted in Figure S6.

The magnitude of measured charge transferred from the solid to the liquid is dependent upon the contact area of the droplet [40]. To better understand what proportion of deposited surface charge is picked up by the droplet during the first wetting cycle, we used the side-profile videos to calculate the change in droplet contact area $\Delta A = A_1 - A_0$, where $A_0$ is the initial contact area upon droplet deposition and $A_1$ is the contact area after the initial wetting. This allowed us to quantify the total deposited surface charge in the wetted area during the first wetting cycle $\Delta q = \sigma_d \cdot \Delta A$, shown in Figure 3 e. The dashed line indicates a 1:1 relationship, demonstrating that ~100 % of the deposited surface ions encountered by the drop are picked up during wetting, regardless of the properties of the surface. The change of contact area on PTFE and Nylon surfaces during each individual wetting and dewetting cycle is plotted in Figure S7, with no significant difference observed across wetting/dewetting stages.

The increased scatter in Figure 3 e. at high deposited surface charge could potentially be due to increased droplet movement on the surface, meaning that the droplet encounters and picks up more surface charge as it wets more of the surface. To further investigate this, we stacked each frame from the side-view video of the first two wetting and dewetting cycles into a composite image. From the composite image, we then measured the total contact line displacement (see Figure S8 for details) and plotted this as a function of deposited charge density (Figure 4 a and b). The contact line displacement increases linearly with the deposited surface charge density for both positive and negative charging on both PTFE and Nylon surfaces (Figure 4 a and b). The PTFE surface shows greater contact line displacement than the Nylon surface, because it is more hydrophobic than Nylon and the contact line is more likely to be shifted during the wetting stage. Reconsidering, the previous observations of increased scatter in the transferred charge data at high initial surface charges can be attributed to increased droplet mobility.

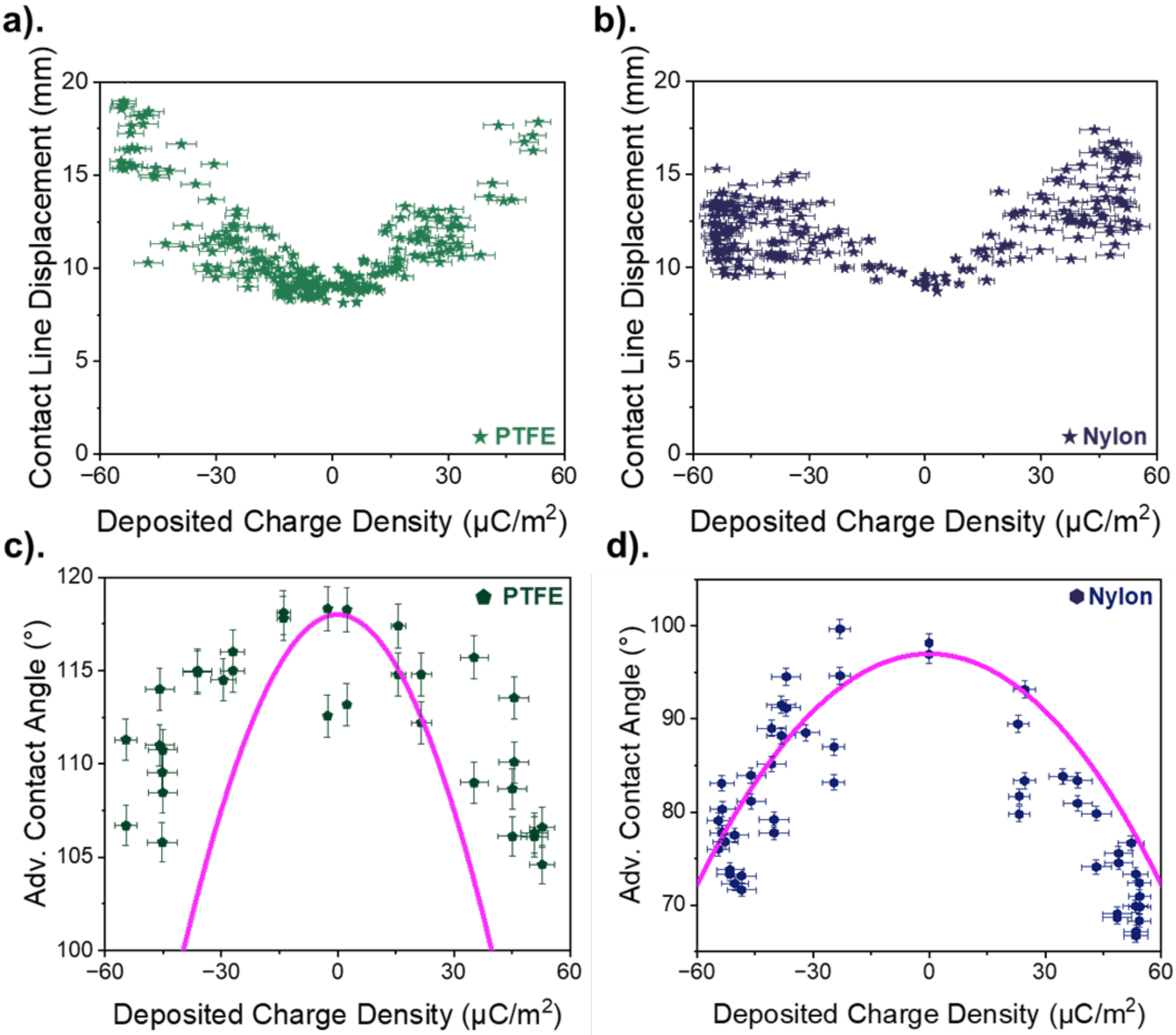

Figure 4. Corresponding contact line displacement (blue squares) plotted against the magnitude of deposited surface charge density on **a.** PTFE surface; **b.** Nylon surface; Corresponding contact angle measurement (purple triangle) plotted against the magnitude of deposited surface charge density on **c.** PTFE surface; **d.** Nylon surface. The magenta line represents the predicted contact angle based on Equation 2.

We also determined the contact angles during the first wetting stage from the side-view videos. For contact angle measurements on both PTFE and nylon surfaces, the advancing contact angle measured was inversely proportional to the magnitude of the deposited surface charge density, regardless of the polarity of the surface charge (Fig. 4 c,d). One possible explanation is an increase of the effective surface energy of the solid as suggested by Li et al. [25] and Moon et al. [41]. The presence of surface charge on the free solid surface increases the effective solid surface energy proportional to $\sigma_d^2$. Assuming a one-dimensional insulating layer of thickness $d$ (which extends infinitely in the other two directions) and with a dielectric permittivity $\varepsilon_d$ which is placed on a grounded plate and with an electrode in air far above the layer, the increase in effective solid surface energy due to homogeneous surface charge can be expressed as [41] $\Delta\gamma_S = \frac{\sigma_d^2 d}{2\varepsilon_d \varepsilon_0}$. Here, $\varepsilon_d$ is the permittivity of the dielectric material, $\varepsilon_0$ is the permittivity of vacuum ($8.854\times10^{-12}$ F·m$^{-1}$), and $d$ is the thickness of dielectric material. Using Young's equation and assuming local a balance of force acting on the contact line the change in contact angle is [26]:

$$cos\theta = cos\theta_0 + \frac{\sigma_d^2 d}{2\gamma\varepsilon_0\varepsilon_d} \qquad \text{Equation 2}$$

where $\theta_0$ is the contact angle on a neutralised surface and $\gamma$ is the surface tension of the liquid. The calculated results (magenta lines in Fig. 4 c,d) show good agreement with the measurements on both the PTFE and Nylon surfaces, supporting the explanation that an increase in deposited surface charge increases the surface energy and leads to enhanced droplet mobility.

**4. Conclusion**

Controlled pre-charging of dielectric polymer surfaces (PTFE and Nylon) systematically modifies solid–liquid charge transfer and droplet behaviour. Using neutralisation and ion-gun pre-charging together with sessile-drop charge measurements across an extensive dataset (>500 trials), we show that droplets pick up pre-deposited surface ions during the first wetting event, with the transferred charge directly correlated to the deposited charge encountered by the wetted area (100% for deposited surface charge density $|\boldsymbol{\sigma_d}| \lesssim \mathbf{40}\ \boldsymbol{\mu}$C/m$^2$) independent of material choice.

We also show that deposited charge alters wetting behaviour, with a decrease in advancing contact angle and corresponding increase in contact-line mobility. This is consistent with an increase in effective solid surface energy due to the presence of surface charge. Higher deposited surface charge magnitudes also initiated instabilities such as droplet splitting or detachment, revealing a trade-off between increasing interfacial charging and maintaining droplet stability.

These findings reveal a previously unrecognised charge-transfer pathway at the solid–liquid interface, which may be exploited to amplify or suppress surface charging and to control fluid motion via electrical input in a wide range of practical applications.

**Acknowledgements**

PCS acknowledges support from RMIT University *via* the RMIT Vice-Chancellors' Fellowship Scheme (2023). JDB is the recipient of an Australian Research Council Future Fellowship (FT220100319) funded by the Australian Government.

**References**

1. Lenard, P., *Ueber die Electricität der Wasserfälle.* Annalen der Physik, 1892. **282**(8): p. 584–636.
2. Thomson, J., *XXXI. on the electricity of drops.* The London, Edinburgh, and Dublin Philosophical Magazine and Journal of Science, 1894. **37**(227): p. 341–358.
3. Yatsuzuka, K., Y. Mizuno, and K. Asano, *Electrification phenomena of pure water droplets dripping and sliding on a polymer surface.* Journal of electrostatics, 1994. **32**(2): p. 157–171.
4. Shahzad, A., K.R. Wijewardhana, and J.-K. Song, *Contact electrification efficiency dependence on surface energy at the water-solid interface.* Applied Physics Letters, 2018. **113**(2).
5. Stetten, A.Z., et al., *Slide electrification: charging of surfaces by moving water drops.* Soft Matter, 2019. **15**(43): p. 8667–8679.
6. Sun, Q., et al., *Surface charge printing for programmed droplet transport.* Nature materials, 2019. **18**(9): p. 936–941.
7. Zhang, W., et al., *Surface charges as a versatile platform for emerging applications.* Science bulletin, 2020. **65**(13): p. 1052–1054.
8. Lin, S., et al., *Quantifying electron-transfer in liquid-solid contact electrification and the formation of electric double-layer.* Nature communications, 2020. **11**(1): p. 399.
9. Helseth, L.E., *A water droplet-powered sensor based on charge transfer to a flow-through front surface electrode.* Nano Energy, 2020. **73**: p. 104809.
10. Zhang, Y., et al., *Rotational electromagnetic energy harvester for human motion application at low frequency.* Applied Physics Letters, 2020. **116**(5).
11. Ratschow, A.D., et al., *How Charges Separate when Surfaces Are Dewetted.* Physical Review Letters, 2024. **132**(22).
12. Sun, Y., X. Huang, and S. Soh, *Using the gravitational energy of water to generate power by separation of charge at interfaces.* Chemical science, 2015. **6**(6): p. 3347–3353.
13. Yatsuzuka, K., Y. Higashiyama, and K. Asano, *Electrification of polymer surface caused by sliding ultrapure water.* IEEE Transactions on Industry Applications, 1996. **32**(4): p. 825–831.
14. Ratschow, A.D., et al., *Liquid slide electrification: advances and open questions.* Soft Matter, 2025.
15. Weidenhammer, P. and H.-J. Jacobasch, *Investigation of adhesion properties of polymer materials by atomic force microscopy and zeta potential measurements.* Journal of Colloid and Interface Science, 1996. **180**(1): p. 232–236.
16. Marinova, K., et al., *Charging of oil− water interfaces due to spontaneous adsorption of hydroxyl ions.* Langmuir, 1996. **12**(8): p. 2045–2051.
17. Zimmermann, R., S. Dukhin, and C. Werner, *Electrokinetic measurements reveal interfacial charge at polymer films caused by simple electrolyte ions.* The Journal of Physical Chemistry B, 2001. **105**(36): p. 8544–8549.
18. Beattie, J.K. and A.M. Djerdjev, *The pristine oil/water interface: Surfactant-free hydroxide-charged emulsions.* Angewandte Chemie International Edition, 2004. **43**(27): p. 3568–3571.
19. Kudin, K.N. and R. Car, *Why are water− hydrophobic interfaces charged?* Journal of the American Chemical Society, 2008. **130**(12): p. 3915–3919.
20. McCarty, L.S. and G.M. Whitesides, *Electrostatic Charging Due to Separation of Ions at Interfaces: Contact Electrification of Ionic Electrets.* Angewandte Chemie International Edition, 2008. **47**(12): p. 2188–2207.

21. Li, X., et al., *Surfactants Screen Slide Electrification.* Angew Chem Int Ed Engl, 2025. **64**(31): p. e202423474.
22. Helseth, L., *Influence of Surface-Active Molecules in Solution on Charge Transfer Due to a Water–Air Contact Line Moving over a Hydrophobic Surface.* Langmuir, 2025. **41**(15): p. 9716–9728.
23. Zhou, X., et al., *Spontaneous Charging from Sliding Water Drops Determines the Interfacial Deposition of Charged Solutes.* Advanced Materials, 2025: p. 2420263.
24. Li, X., et al., *Spontaneous charging affects the motion of sliding drops.* Nature Physics, 2022. **18**(6): p. 713–719.
25. Li, X., et al., *Surface Charge Deposition by Moving Drops Reduces Contact Angles.* Physical Review Letters, 2023. **131**(22).
26. Hinduja, C., et al., *How Spontaneous Electrowetting and Surface Charge affect Drop Motion.* arXiv preprint arXiv:2602.03362, 2026.
27. Monluc, H., et al., *Impact of Initial Charge Conditions on the Slide Electrification of Droplets.* Langmuir, 2025. **41**(40): p. 27227–27238.
28. Hinduja, C., H.-J. Butt, and R. Berger, *Slide Electrification of Drops at Low Velocities.* Soft Matter, 2024.
29. Gao, N., et al., *How drops start sliding over solid surfaces.* Nature Physics, 2018. **14**(2): p. 191–196.
30. Chen, S., et al., *Irreversible Charging Caused by Energy Dissipation from Depinning of Droplets on Polymer Surfaces.* Physical Review Letters, 2025. **134**(10): p. 104002.
31. Lin, S., X. Chen, and Z.L. Wang, *The tribovoltaic effect and electron transfer at a liquid-semiconductor interface.* Nano Energy, 2020. **76**: p. 105070.
32. Chen, K.T., et al., *Repulsion, Acceleration, and Coalescence between Water Droplets on Superhydrophobic Glass by Triboelectrification.* Langmuir, 2024. **40**(25): p. 13219–13226.
33. Wu, H., et al., *Energy Harvesting from Drops Impacting onto Charged Surfaces.* Physical Review Letters, 2020. **125**(7).
34. Lacks, D.J. and T. Shinbrot, *Long-standing and unresolved issues in triboelectric charging.* Nature Reviews Chemistry, 2019. **3**(8): p. 465–476.
35. Burch, P., *Cosmic radiation: ionization intensity and specific ionization in air at sea level.* Proceedings of the Physical Society. Section A, 1954. **67**(5): p. 421–430.
36. Bista, P., et al., *High voltages in sliding water drops.* The journal of physical chemistry letters, 2023. **14**(49): p. 11110–11116.
37. Ratschow, A.D., et al., *How charges separate when surfaces are dewetted.* arXiv preprint arXiv:2305.02172, 2023.
38. Taflin, D.C., T.L. Ward, and E.J. Davis, *Electrified droplet fission and the Rayleigh limit.* Langmuir, 1989. **5**(2): p. 376–384.
39. Lin, M., et al., *Spontaneous Coulomb fissions of drops on lubricated surfaces.* arXiv preprint arXiv:2510.10368, 2025.
40. Adhikari, P.R., et al., *High power density and bias-free reverse electrowetting energy harvesting using surface area enhanced porous electrodes.* Journal of Power Sources, 2022. **517**: p. 230726.
41. Moon, H., et al., *Low voltage electrowetting-on-dielectric.* Journal of applied physics, 2002. **92**(7): p. 4080–4087.

**Supporting Information**

**Pre-charging polymer surfaces enhances droplet mobility and electrification**

*Shuaijia Chen[1], Kenta Morita[2,3], Dumindu Dassanayaka[4], Hans-Jürgen Butt[5], Peter C. Sherrell[1,2],* Amanda V. Ellis[1],* Joseph D. Berry[1]**

S. Chen, K. Morita, D. Dassanayaka, H.J. Butt, P. C. Sherrell, A. V. Ellis, J. D. Berry,
Department of Chemical Engineering, The University of Melbourne, Parkville, 3010, Victoria Australia
E-mail: amanda.ellis@unimelb.edu.au ; berryj@unimelb.edu.au

P. C. Sherrell,
School of Science, RMIT University, Melbourne, 3000, Victoria, Australia
Email: peter.sherrell@rmit.edu.au

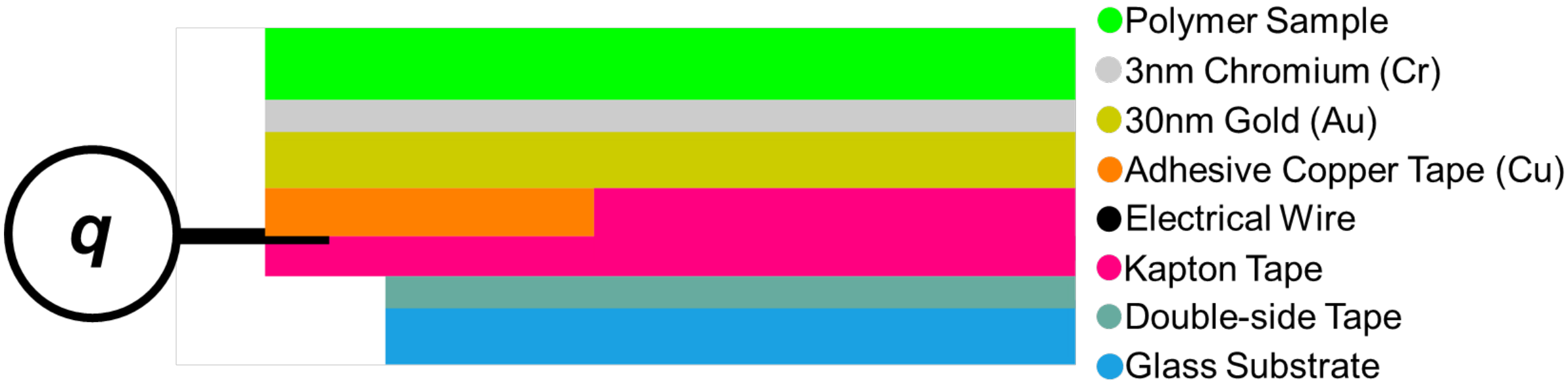

**Figure S1.** Schematic of the testing polymer sample and connection to the electrometer

## 1. Metrology of pre-charging

During the experiment, we noticed that directly charging the surface with the anti-static gun as non-shielded charging (Figure S1 a) could result in charging the attached electric wire, resulting in a much higher reading from the electrometer than the actual value of deposited surface charge density, which leads to a misleading result, and it was not consistent with other literature. To fix the experimental protocol and existing results, we add a 3D printed channel to hold the anti-static gun as the shielded charging protocol (Figure S1 b) to ensure consistent pre-charging. After that, we applied the linear fitting to the results with the shielded charging protocol to establish a fix factor for both PTFE and Nylon surfaces. By applying this fix factor to the charge transfer results from the non-shielded charging protocol on PTFE surface during the first wetting stage, the results from both protocols show a good alignment (Figure S1 c and d), then the fix factor is applied to all the results in the manuscript.

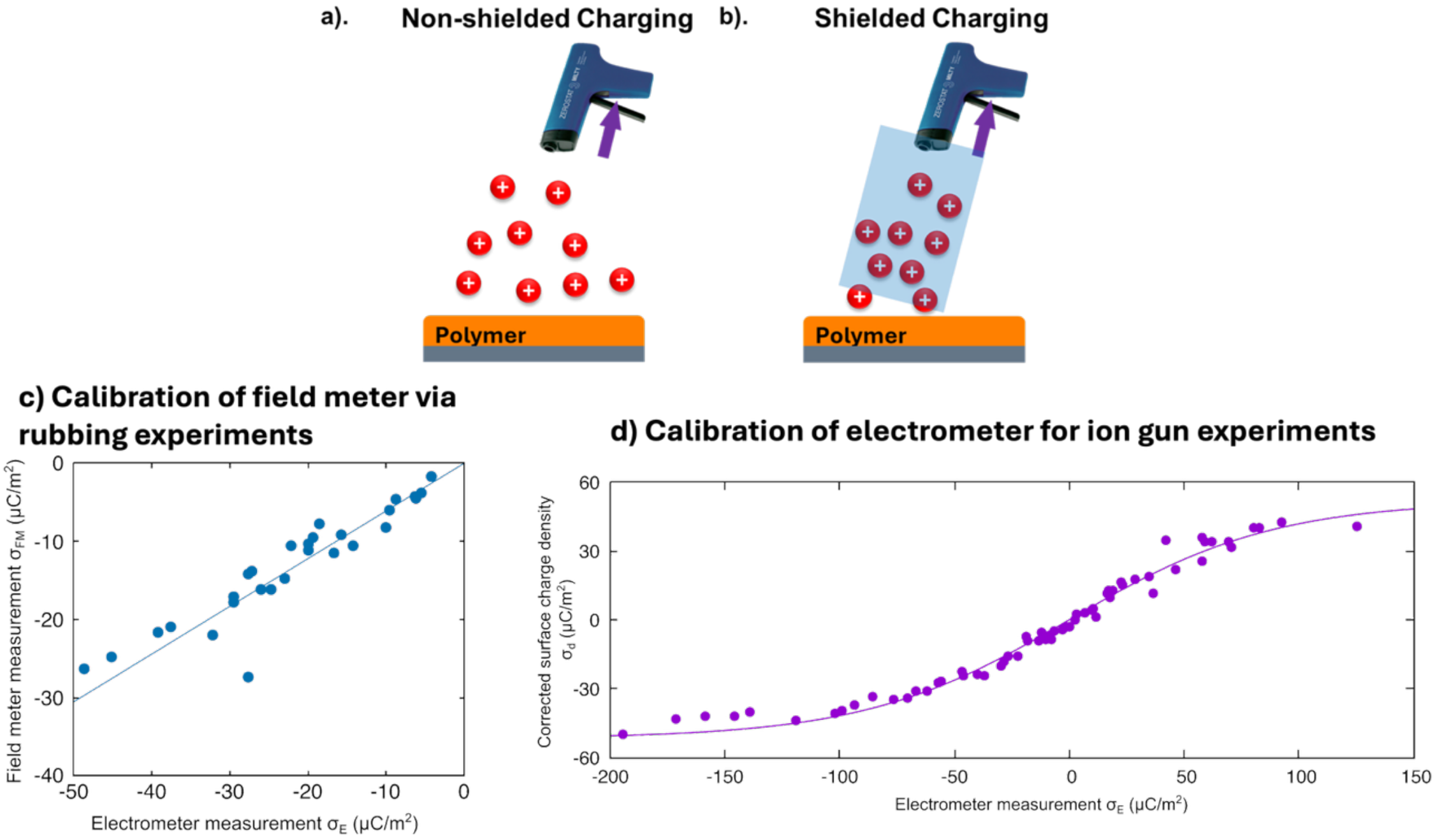

**Figure S2.** Schematic of **a.** non-shielded pre-charging protocol; **b.** shielded pre-charging protocol (blue shaded region is the 3D-printed channel **c.** calibration of the field meter data using surfaces charged via rubbing (ie without charge leakage); **d.** calibration of electrometer estimates of surface charge density for ion-gun charging that eliminates charge leakage.

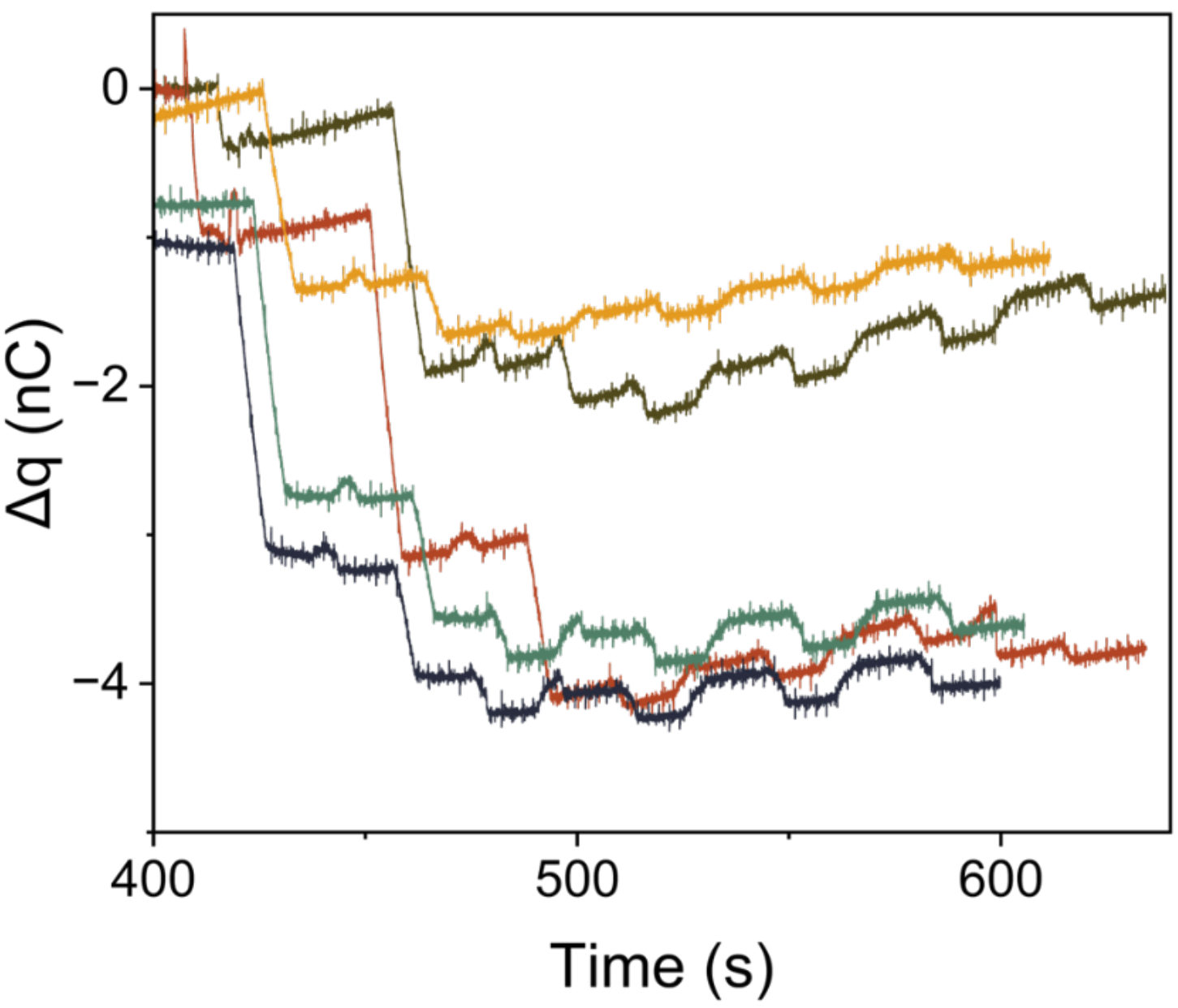

**Figure S3.** Examples of charge measurement with 5 cyclic wetting and dewetting stages on a positively pre-charged PTFE surface. After the first two cycles of wetting and dewetting, the charge transfer Δq becomes constant in the subsequent cycle. Therefore, the results in the manuscript mainly focus on the charge transfer during the first two cycles of wetting and dewetting.

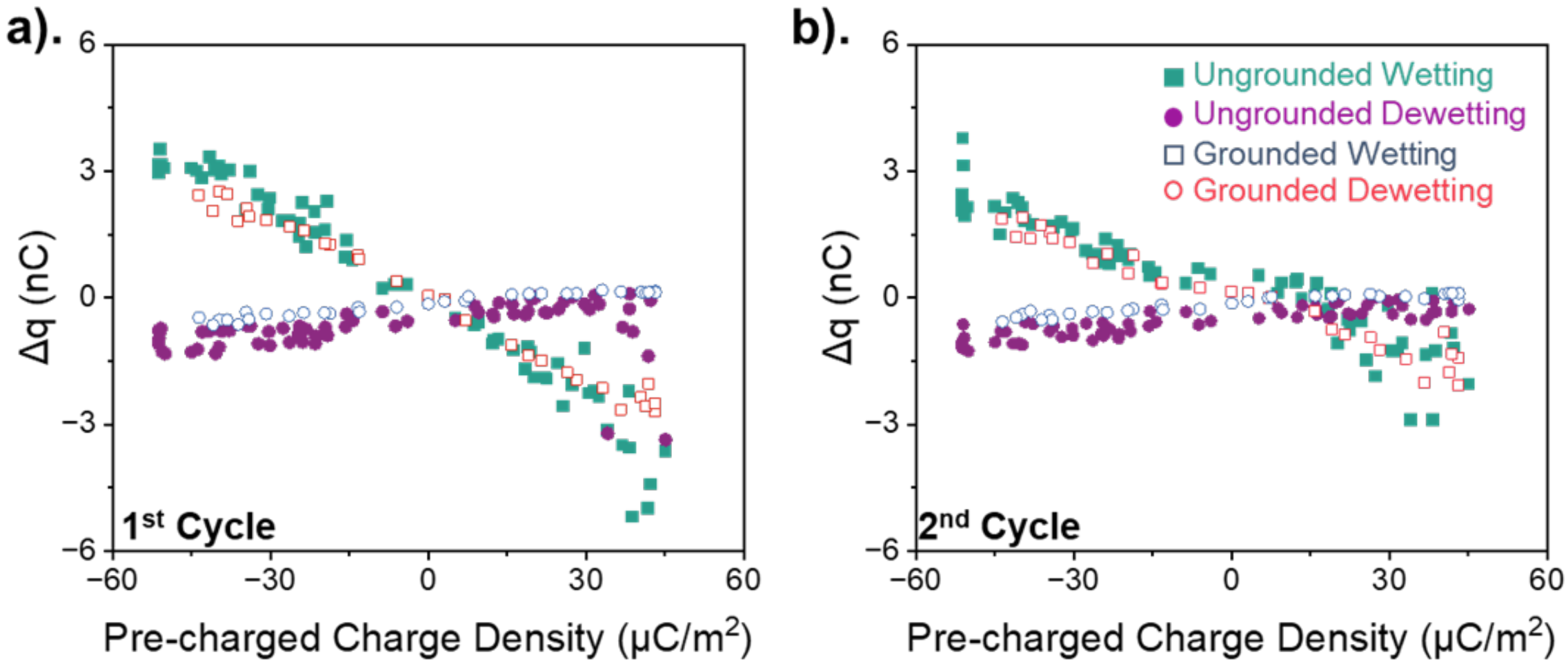

**Figure S4.** The comparison of the grounding needle and ungrounded needle protocols on PTFE surface plotted against the magnitude of deposited surface charge density during the **a.** first cycle; **b.** second cycle

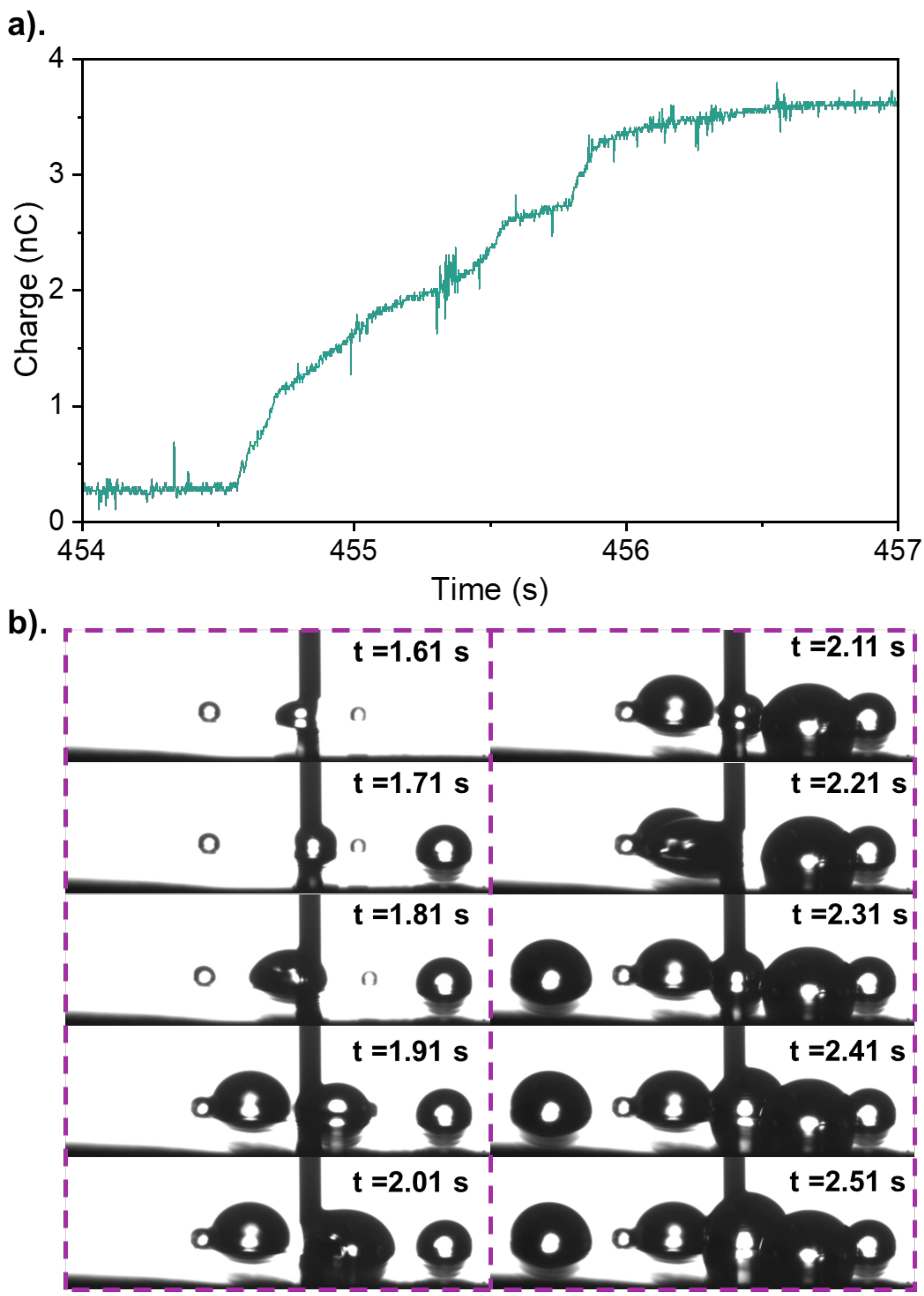

**Figure S5. a.** Charge measurement during the droplet splitting event with a magnitude of pre-charging surface charge density at saturation; **b.** extracted sequence of droplet images during the splitting event.

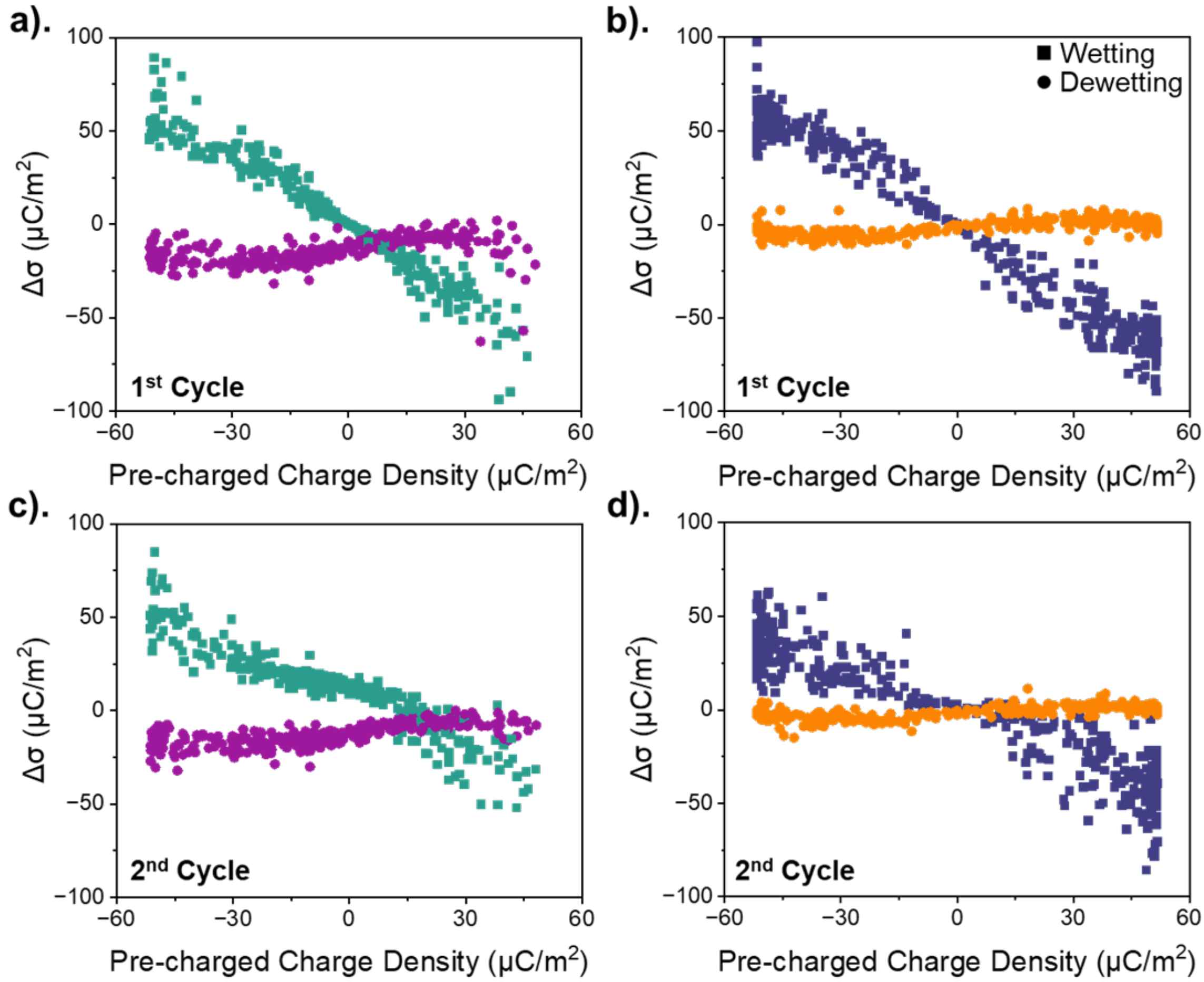

**Figure S6.** The comparison of change of charge density between wetting and dewetting plotted against the magnitude of deposited surface charge density during the first cycle on **a.** PTFE surface; **b.** Nylon surface; the second cycle on **c.** PTFE surface; **d.** Nylon surface

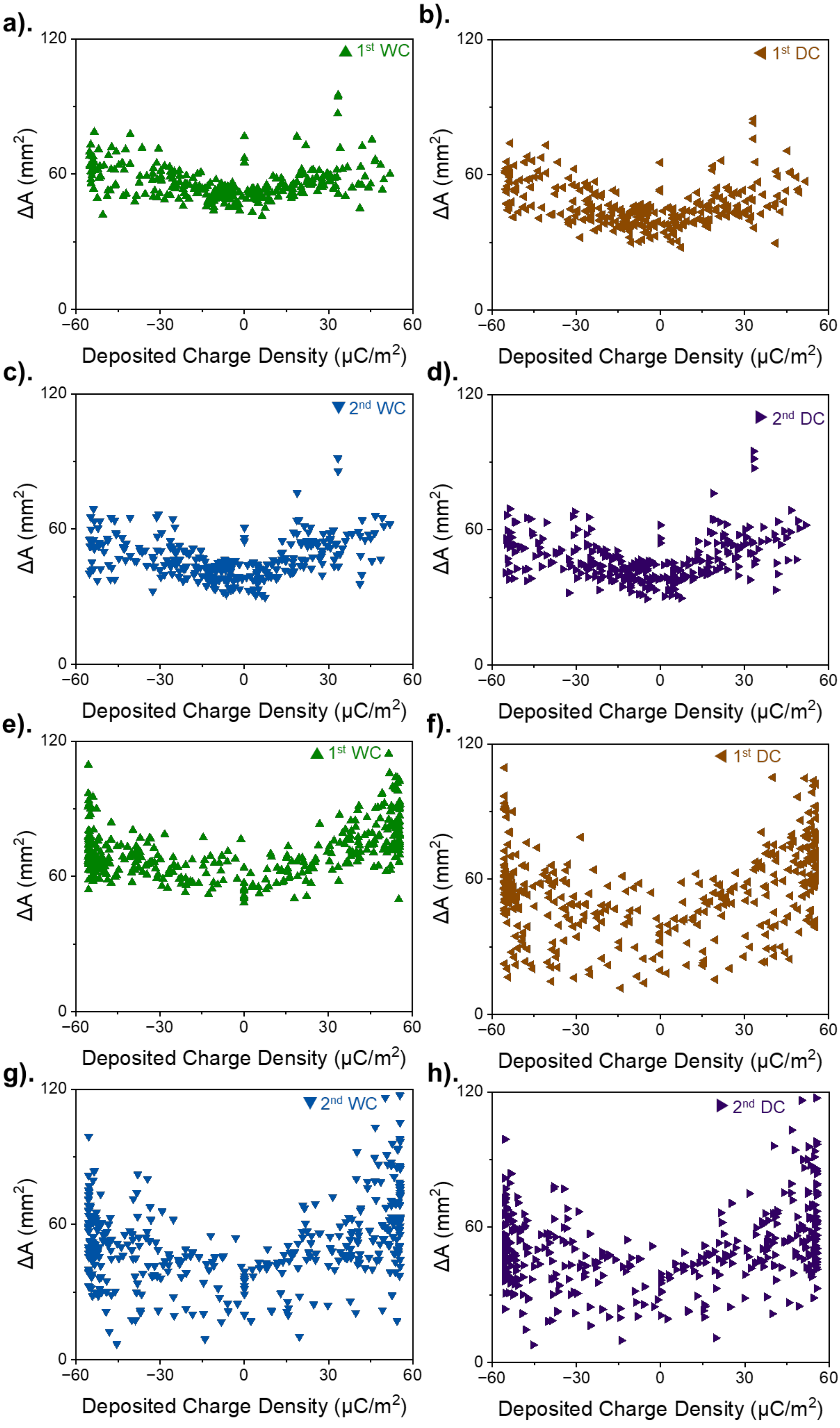
a).
b).
c).
d).
e).
f).
g).
h).
1st WC
1st DC
2nd WC
2nd DC
ΔA (mm2)
Deposited Charge Density (μC/m2)
120
60
0
−60
−30
0
30
60

**Figure S7.** The change of contact area plotted against the magnitude of deposited surface charge density on PTFE surface during **a.** the first wetting cycle; **b.** the first dewetting cycle; on **c.** the second wetting cycle; **d.** the second dewetting cycle; on Nylon surface during **e.** the first wetting cycle; **f.** the first dewetting cycle; on **g.** the second wetting cycle; **h.** the second dewetting cycle.

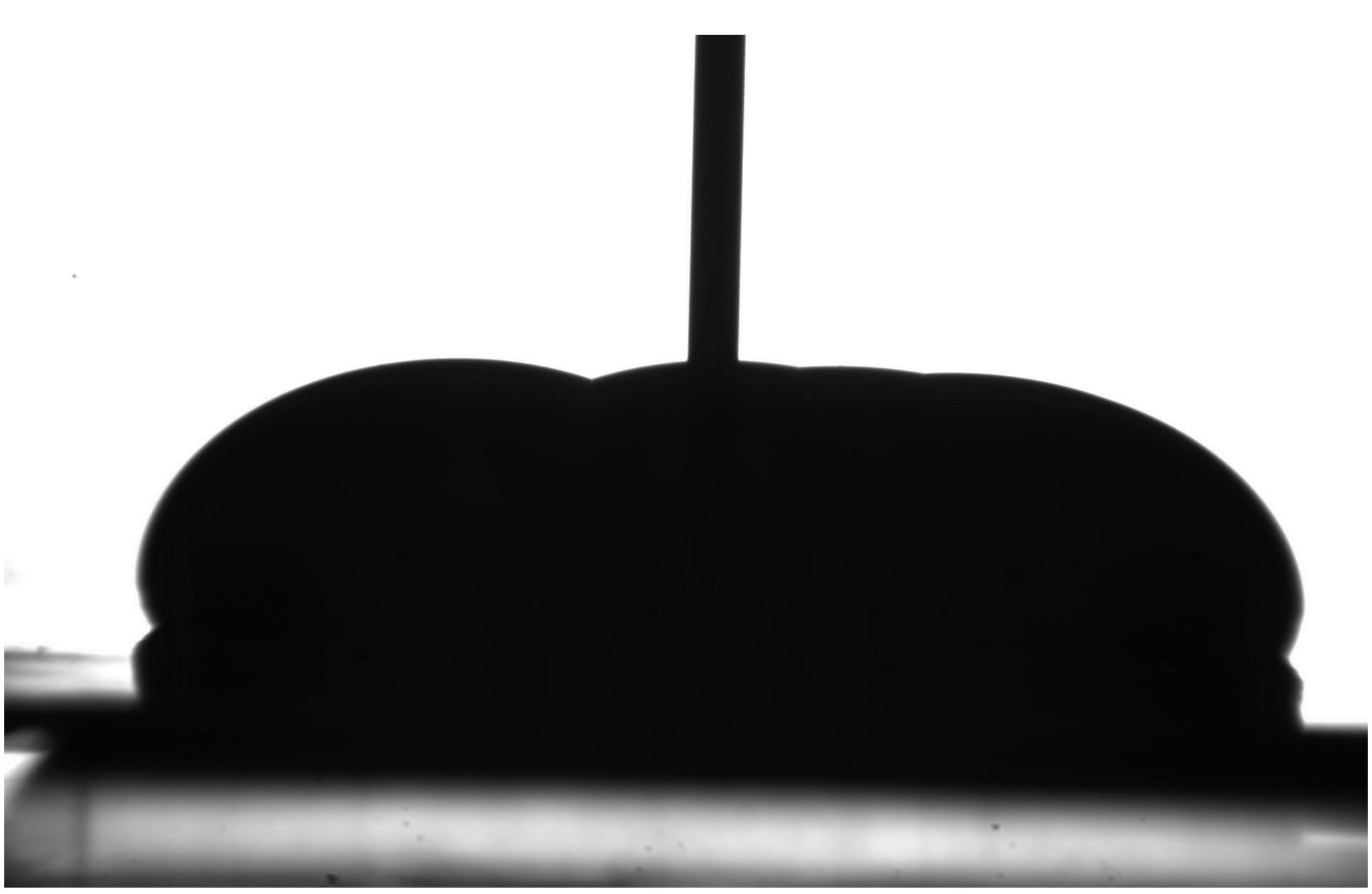

**Figure S8.** Example of a composite image generated from two wetting/dewetting cycles of the droplet on PTFE. The total contact line displacement is measured from this image and plotted in Figure 4.

## 2. Calculation of droplet stability and splitting threshold

To evaluates the critical value of charge density causing droplet splitting, following calculation is performed based on the theory of Rayleigh Limit [1].

The stability of the liquid drop can be described as:

$$q^2 < X64\pi^2\varepsilon_0\gamma a^3 \qquad \text{Equation 1}$$

Where $q$ is the charge, $\varepsilon_0$ is the permittivity of free space, $\gamma$ is the surface tension, and a is the radius of the liquid drop, and X is the fissility threshold [2] .

The drop volume is:

$$V = \frac{4}{3}\pi a^3 \Rightarrow a^3 = \frac{3V}{4\pi} \qquad \text{Equation 2}$$

Substitute this into Equation 1:

$$q^2 < 48\pi\varepsilon X_0\gamma V \qquad \text{Equation 3}$$

Assume $\theta = 90°$, then:

$$V = \frac{2}{3}\pi R_c^3 \Rightarrow R_c^3 = \frac{3V}{2\pi} \Rightarrow R_c^2 = \left(\frac{3V}{2\pi}\right)^{\frac{2}{3}} \qquad \text{Equation 4}$$

Where $R_c$ is the contact radius.

The contact area is represented by:

$$A_c = \pi R_c^2 \qquad \text{Equation 5}$$

The surface charge available for the droplet to interact with via its contact area is represented by:

$$q = \sigma_s A_c = \pi\sigma_s R_c^2 \Rightarrow q = \pi\sigma_s \left(\frac{3V}{2\pi}\right)^{\frac{2}{3}} \Rightarrow q^2 = \pi^2\sigma_3^2 \left(\frac{3V}{2\pi}\right)^{\frac{4}{3}} \qquad \text{Equation 6}$$

Substitute this into Equation 6:

$$\pi^2\sigma_s^2 \left(\frac{3V}{2\pi}\right)^{\frac{4}{3}} < 48\pi X\varepsilon_0\gamma V \qquad \text{Equation 7}$$

$$\sigma_s^2 < \frac{48X\varepsilon_0\gamma}{\pi}\left(\frac{2\pi}{3V}\right)^{\frac{4}{3}} V \qquad \text{Equation 8}$$

$$\sigma_s^2 < \frac{48X\varepsilon_0\gamma}{\pi}\left(\frac{2\pi}{3}\right)^{\frac{4}{3}} V^{-\frac{1}{3}} \qquad \text{Equation 9}$$

$$\sigma_s < \left(\frac{48X\varepsilon_0\gamma}{\pi}\left(\frac{2\pi}{3}\right)^{\frac{4}{3}}\right)^{\frac{1}{2}} V^{-\frac{1}{6}} \qquad \text{Equation 10}$$

Therefore, the stability of liquid drop brought into contact with charged dielectric surface can be represented by Equation 10 based on the droplet volume.

1. Taflin, D.C., T.L. Ward, and E.J. Davis, *Electrified droplet fission and the Rayleigh limit.* Langmuir, 1989. **5**(2): p. 376–384.
2. Lin, M., et al., *Spontaneous Coulomb fissions of drops on lubricated surfaces.* arXiv preprint arXiv:2510.10368, 2025.